\documentclass[twocolumn]{aastex63}

\usepackage{xcolor}

\usepackage{bm}
\usepackage{amsmath}
\usepackage{empheq}
\usepackage{enumitem}

\accepted{15-Oct-2021}

\newcommand{\esp}[1]{\mathbb{E}\left( {#1} \right)}
\newcommand{\var}[1]{\mathrm{Var}\left( {#1} \right)}

\defcitealias{jaupart2020}{JC20}
\defcitealias{Jaupart21}{JC21}

\shorttitle{Density PDFs}
\shortauthors{Jaupart \& Chabrier}
\graphicspath{{./}{figures/}}

\begin{document}

\title{Generalised transport equation of the Autocovariance Function of the density field and mass invariant  in star-forming clouds}



\author{Etienne Jaupart}
\affiliation{Ecole normale sup\'erieure de Lyon, CRAL, Universit\'e de Lyon, UMR CNRS 5574, F-69364 Lyon Cedex 07, France}

\author{Gilles Chabrier}
\affiliation{Ecole normale sup\'erieure de Lyon, CRAL, Universit\'e de  Lyon, UMR CNRS 5574, F-69364 Lyon Cedex 07, France}
\affiliation{School of Physics, University of Exeter, Exeter, EX4 4QL, UK}

\email{etienne.jaupart@ens-lyon.fr}

\begin{abstract}

In this Letter, we study the evolution of the autocovariance function (ACF) of density field fluctuations in star-forming clouds and thus of the correlation length $l_c(\rho)$ of these fluctuations, which can be identified as the average size of the most correlated structures within the cloud. Generalizing the transport equation
derived by \citet{chandra1951} for static, homogeneous turbulence, we show that the mass contained within these structures is an invariant, i.e. that the average mass contained in the most correlated structures remains constant during the evolution of the cloud, whatever dominates the global dynamics (gravity or turbulence).
  We show that the growing impact of gravity on the turbulent flow yields an increase of the variance of the
density fluctuations and thus a drastic decrease of the correlation length. Theoretical relations are successfully compared to numerical simulations.
This picture  brings a robust support to star formation paradigms where the mass concentration in turbulent star-forming clouds evolves from initially large, weakly correlated filamentary structures to smaller, denser more correlated ones,  and eventually to small, tightly correlated prestellar cores. We stress that the present results rely on a pure statistical approach of density fluctuations and do not involve any specific
condition for the formation of prestellar cores.
Interestingly enough, we show that, under average conditions typical of Milky Way molecular clouds, this invariant average mass is about a solar mass, providing an appealing explanation for the apparent universality of the IMF under such environments.

\end{abstract}



\section{Introduction} \label{sec:intro}

The dynamics of star-forming molecular clouds (MCs) is determined by the statistical properties of their density fluctuations, under the action of turbulence and gravity. A fundamental quantity in such a study
is the autocovariance function (ACF) of density field fluctuations, which allows the determination of the characteristic correlation length $l_c(\rho)$ of density structures within the cloud. In this Letter, we study the ACF and the correlation length of density fluctuations in MCs and we show that this latter can be identified as the average size of the most correlated structures within the cloud. Generalizing the transport equation
derived by \citet{chandra1951} for static, homogeneous isotropic turbulence to a non-isotropic, time evolving turbulent flow, we show that, whereas the correlation length
decreases with time as gravity proceeds in the cloud, the mass contained within these structures of size $l_c(\rho)$ is an invariant,
like invariants found e.g. in incompressible turbulence \citep{batchelor1953}. This striking result implies that the average mass contained in the most correlated structures in star-forming gravo-turbulent MCs, which will be ultimately distributed within prestellar cores, is imprinted within the initial conditions of the cloud and is constant during its evolution.  

\section{Mathematical Framework} \label{sec:mathframe}
\subsection{Evolution of a molecular cloud}
The dynamics of the cloud is described as in \citet{jaupart2020} (hereafter \citetalias{jaupart2020}). The only useful equation for the present study is mass conservation:
\begin{eqnarray}
    \frac{\partial \rho}{\partial t} + \bm{\nabla}\cdot \left( \rho \bm{v} \right) &=& 0, \label{eq::masscons} 
\end{eqnarray}
where $\rho$ denotes the gas mass density and $\bm{v}$ the velocity field 


We are interested in clouds that will eventually condense locally to form stars, and hence we separate the evolution of the background from that of local density deviations. 
The velocity field $\bm{v}$ is thus split into a mean velocity $\bm{V}$ and a (turbulent) velocity $\bm{u}$  (\citealt{ledoux1958}). Introducing the logarithmic excess of density,  $s=\log(\rho/\overline{\rho})$, we get by definition: 
\begin{eqnarray}
    \bm{V} &\equiv& \frac{1}{\overline{\rho}} \overline{\rho \bm{v}} \label{eq::meanV}, \\
    \bm{u} &\equiv& \bm{v} - \bm{V}, \\
    \rho &\equiv& \overline{\rho}(\bm{x},t) \, e^s, \label{eq:defs}
\end{eqnarray}
where $\overline{\Phi}(\bm{x},t) \equiv \esp{\Phi}(\bm{x},t)$ is the mathematical expectation, also called statistical average or mean, of  random field $\Phi$ (\citealt{pope1985,frisch1995}). 
We note that  $\overline{\bm{u}} \neq 0$ \textit{a priori} but $\overline{\rho \bm{u}} = 0$. This ensures that on average there is no transfer of mass due to turbulence 
and the equation of continuity (\ref{eq::masscons}) remains valid for the mean field,
\begin{eqnarray}
    \frac{\partial \overline{\rho}}{\partial t} + \bm{\nabla}\cdot \left( \overline{\rho} \ \bm{V}  \right) &=& 0. \label{eq::massconsav}
\end{eqnarray}

Subtracting the equations for the average variables from the original equations, we obtain the evolution of the density deviations. 

\subsection{Model for the statistics of a turbulent cloud}\label{subsec:modelizationturb}
\subsubsection{Statistically homogeneous clouds} \label{subsec:classofflows}

In studies of star formation, be it observations of a cloud or numerical simulations, one has usually access to only a small number of samples (only one in most cases). 
Thus, one has to make the basic assumption, sometimes called "fair-sample hypothesis", that the observed sample is large enough 
for volumetric (or time) averages over this single sample to provide accurate statistical estimates. For this procedure to be valid, 
the random field \textit{must} be ergodic and thus \textit{statistically} homogeneous \citep{papoulis1965}. Note that \textit{statistical} homogeneity does not imply \textit{spatial} homogeneity. 
 Ergodicity, one of the fundamental hypothesis of statistical physics, insures that the average value of a statistical quantity (density fluctuations in the present context) is equal to the mean of a large number (in space or time) of measured quantities
 (e.g. \citealt{penrose1979foundations}).
It is commonly made for instance in studies of turbulent flows, with or without self gravity \citep{chandra1951,chandra1951gravity,batchelor1953,pope1985,frisch1995,pan2018,Pan2019A,Pan2019B,jaupart2020} 
or in cosmology to study the dynamical evolution of structures in the Universe \citep{Peebles1973,heinesen2020}. This assumption does not constrain fluctuations around the average to be small. 
Statistical homogeneity implies that, for any stochastic field $\Phi$, $\overline{\Phi(\bm{x},t)} = \overline{\Phi}(t)$. In particular, $\overline{\rho}(\bm{x},t) = \overline{\rho}(t)$ in our context. 

With these assumptions, the dynamics of the cloud density and logarithm of density fluctuations are governed by the following equations:
\begin{eqnarray}
    - \frac{\mathrm{d} \mathrm{ln}(\overline{\rho})}{\mathrm{d} t} &=& - \frac{1}{\overline{\rho}} \frac{\mathrm{d} \overline{\rho}}{\mathrm{d} t} = \bm{\nabla} \cdot \bm{V}, \label{eq::divVhomostrict}\\
      \frac{\mathrm{D} s}{\mathrm{D} t} &=& - \bm{\nabla} \cdot \bm{u}, \label{eq::advectshomostrict} 
\end{eqnarray}
where $\frac{\mathrm{d}}{\mathrm{d}t}$ denotes the derivative of a variable that is only a function of time $t$ and $\frac{\mathrm{D}}{\mathrm{D} t}=\frac{\partial}{\partial t} + \left(\bm{v}\cdot \bm{\nabla}\right)$ is the Lagrangian derivative.
Eq.~(\ref{eq::divVhomostrict}) shows that the statistical homogeneity hypothesis for $\rho$ implies, to be consistent, that the r.h.s. of Eq.(\ref{eq::divVhomostrict}) must be a function of time $t$ only. It thus constrains the flow to belong to a certain class of flows. 
In order to fullfill this constraint, it suffices that :
\begin{empheq}[box=\fbox]{align}
\bm{V}(\bm{x},t) = \underline{\underline{\mathrm{L}_V}} (t) \cdot \bm{x} + \bm{c}_V(t) \label{eq:generalformV},
\end{empheq}
where $ \underline{\underline{\mathrm{L}_V}} (t)$ is a $3\times3$ matrix and $\bm{c}_V(t)$ is a spatially constant vector.
Enforcing $\bm{V}=0$ yields exactly the  equations  usually used to prescribe the evolution of a periodic simulation box  in an astrophysical context \citep{federrath2012,Pan2019B}. 
 However, this is not equivalent to applying periodic boundary conditions (see, e.g., \citealt{robertson2012} for an example 
of periodic box and $\bm{V} \ne 0 $).

\subsubsection{Accepted class of flows}

In our homogeneous model, the bulk flow $\bm{V}$ is restricted to a certain class of flows. This class, however, contains many kinds of flows relevant to the present study, 
such as linearized shears, notably galactic shears, homogeneous rotations, and in particular solid rotations, and global homogeneous contractions or expansions, which need not be isotropic. We note that this construction is similar to that used in Newtonian cosmology, where usually $\bm{V} = H(t)\, \bm{x}$ is the Hubble flow and $H(t)$ is Hubble's expansion rate (see also \citealt{Buchert1997,vigneron2021}  for the class of permitted  flow in cosmology).

Therefore, these models can properly describe the evolution of the density field statistics in star-forming clouds.

\subsection{Ergodicity and the ACF of the homogeneous density field}

In ergodic \textit{theory}, which specifies under which conditions the ergodic \textit{hypothesis} is valid and provides an assessment of errors in the estimation of averages, the  autocovariance function (ACF)  $C_\rho$ of the statistically homogeneous density field is of prime importance (see e.g. \citealt{Jaupart21} hereafter \citetalias{Jaupart21}). It is defined as
\begin{equation}
C_\rho(\bm{x}-  \bm{x'}) \equiv \esp{\rho( \bm{x})\rho( \bm{x'})}- \overline{\rho}(t)^2,
\end{equation}
and reaches a maximum at $\bm{\xi} = \bm{x}-  \bm{x'}=\bm{0}$ : $C_\rho(\bm{\xi}) \leq C_\rho(\bm{0}) = \var{\rho}$  (see e.g. \citealt{papoulis1965} for a demonstration), where $\var{\rho}$ is the variance of $\rho$. Length scales for which correlations are statistically significant  are encoded in the ACF. This statistical object allows thus the extraction of characteristic length scales of physical processes.

\subsubsection{Slutsky's theorem and the correlation length}
As mentioned above, one assumes statistical homogeneity and builds the following ergodic estimator for the expectation of $\rho$:
\begin{equation}
    \hat{\rho}_L = \frac{1}{L^3} \int_{\Omega} \rho(\bm{x}) \, \mathrm{d} \bm{x}, \label{eq:defestimergodic}
\end{equation}
where $\Omega=[-\frac{L}{2},\frac{L}{2}]^3$ is a control volume of linear size L and volume $L^3$, which is sought to be as large as possible. The ergodic estimator $\hat{\rho}_L$ has variance:
\begin{equation}
    \mathrm{Var}( \hat{\rho}_L ) = \frac{1}{L^3} \int_{2 \Omega} C_\rho(\bm{\xi}) \, \prod_{k=1}^3 \left( 1 - \frac{|\xi_k|}{L} \right) \, \mathrm{d} \bm{\xi} \label{eq:varXl},
\end{equation}
where the integration volume $2\Omega=[-{L},+{L}]^3$ stems form the change of variables $(\bm{x},\bm{x'})\rightarrow (\bm{\xi}=\bm{x}-\bm{x'},\bm{y}=\bm{x}+\bm{x'})$.
This leads to Slutsky's theorem \citep{papoulis1965}: the stochastic field $\rho$ is mean ergodic in the mean square (MS) sense, if and only if
\begin{empheq}[box=\fbox]{align}
    \frac{1}{L^3} \int_{2 \Omega} C_\rho(\bm{\xi}) \mathrm{d} \bm{\xi} \xrightarrow[L \rightarrow \infty]{} 0.  \label{eq:Slutsky}
\end{empheq}
From this, one derives two sufficient (physical) conditions for $\rho$ to be mean ergodic. Either:
\begin{equation}
    \int_{\mathbb{R}^3} C_\rho(\bm{\xi}) \mathrm{d} \bm{\xi} < \infty, \label{eq:SufCond1}
\end{equation}
or
\begin{equation}
   C_\rho(\bm{\xi}) \xrightarrow[{| \bm{\xi}| \rightarrow \infty}]{}   0, \label{eq:SufCond2}
\end{equation}
which means that values of the density field at two points separated by a lag $\bm{\xi}$ are uncorrelated at infinitely large distance. The first condition leads to the definition of the correlation length $l_c(\rho)$ of the density field $\rho$ (see e.g. \citealt{papoulis1965}):
\begin{empheq}[box=\fbox]{align}
    (l_c(\rho))^3 = \frac{1}{2^3\,C_\rho(\bm{0})}\int_{\mathbb{R}^3} C_\rho(\bm{\xi})  \mathrm{d} \bm{\xi} = \frac{1}{2^3}\int_{\mathbb{R}^3} \Tilde{C}_\rho(\bm{\xi})  \mathrm{d} \bm{\xi}, \label{eq:defLc}
\end{empheq}
where 
\begin{equation}
\Tilde{C}_\rho(\bm{\xi}) = \frac{C_\rho(\bm{\xi})}{\var{\rho}}
\end{equation}
is the correlation coefficient at lag $\bm{\xi}$ that generates a measure of how correlated two values of the density field  are.
Then, using the two physical assumptions Eqs.~(\ref{eq:SufCond1}) and (\ref{eq:SufCond2}), one obtains for $ l_c(\rho)\ll L$, and from Eq.~(\ref{eq:varXl}):
\begin{equation}
    \mathrm{Var}( \hat{\rho}_L ) \simeq \mathrm{Var}(\rho) \left(\frac{ 2 \, l_c(\rho)}{L}\right)^3 = \mathrm{Var}(\rho) \left(\frac{ l_c(\rho)}{R}\right)^3,
 \label{eq:varergodicestimate}
\end{equation}
where $R=L/2$. Comparing Eq.~(\ref{eq:varergodicestimate}) with the variance  $\var{ \hat{\rho}_{\bm{x},N}}$ of the estimator of the expectation $\overline{\rho}$ obtained from a frequency interpretation where the experiment is repeated over $N$ independent trials $\omega_i$, 
\begin{align}
    \hat{\rho}_{\bm{x},N} = \frac{1}{N} \sum_{i=1}^N \rho(\bm{x}, \omega_i), \label{eq:freqestimate} \\
    \mathrm{Var}(\hat{\rho}_{\bm{x},N})=  \frac{\mathrm{Var}(\rho)}{N} , \label{eq:varfreqestimate}
\end{align}
we see that one can interpret the ratio $(R/l_c(\rho))^3$ as an {\it effective number of "independent" samples}. 

\subsection{The correlation length as an average size of the most correlated structures} \label{sec:corrlengthAvSize}

\subsubsection{Integral scale}
In homogeneous and isotropic turbulence, one introduces a quantity similar to the correlation length, called the integral scale $l_\mathrm{i}$ (not to be confused with the injection scale),
defined as (\citealt{batchelor1953})
\begin{equation}
    l_\mathrm{i}(\rho) = \frac{1}{C_\rho({0})} \int_0^\infty C_\rho(\xi) \mathrm{d}\xi =   \int_0^\infty \Tilde{C}_\rho(\xi) \mathrm{d}\xi \,.
\end{equation}
In the usual phenomenology of turbulence, this integral scale is used as a measure of the lags for which the velocities are significantly correlated and thus gives a measure of the accuracy of volumetric averages as estimates of actual statistical averages \citep{frisch1995}. 
The correlation length, the very quantity that enters Slutsky's theorem (Eq.~\ref{eq:Slutsky}), is given, in this isotropic context, by
\begin{equation}
l_c(\rho)^3 = \frac{\pi}{2} \int_0^\infty C_\rho(\xi) \xi^2 \mathrm{d}\xi.
\end{equation}
One finds that $l_c \simeq l_\mathrm{i}$ in many cases. Indeed, for an exponential ACF  with $C_\rho(\bm{\xi}) = \mathrm{Var}(\rho) e^{-|\bm{\xi}|/l_\mathrm{i}} $, 
we have $l_c=\pi^{1/3} \, l_\mathrm{i}$, whereas for a Gaussian ACF ($C_\rho(\bm{\xi}) = \mathrm{Var}(\rho) e^{-|\bm{\xi}|^2/\lambda}$),  $l_c = l_i$. Moreover, for an ACF of the form $C_\rho(\xi) = \mathrm{Var}(\rho) ( 1-(\xi/l_0)^p)$ for $r < l_0$ and  decaying rapidly outward,   one gets $l_c = (1.9 - 0.8) \,  l_\mathrm{i}$ for $p \in [0.2, \infty[$ (the typical value in turbulence is $p=2/3$ for the velocity field).

The integral scale can thus serve as a proxy for the correlation length, but this latter is the only quantity defined in absence of isotropy, as well as the one entering Slutsky's theorem (Eq.~\ref{eq:Slutsky}).

\subsubsection{Average size of the most correlated structures} \label{subsec:averagesizecorrel}

If the ACF of the ergodic field $\rho$ is isotropic, the above equation for the integral scale $ l_\mathrm{i}(\rho)$ can be used to define a weight function $W_l(\xi)$ that measures the correlation of structures of size $\xi=|\bm{\xi}|$
\begin{equation}
W_l(r)=\frac{1}{ l_\mathrm{i}(\rho)}\frac{C_\rho(r)}{\var{\rho}} = \frac{\Tilde{C}_\rho(r)}{ l_\mathrm{i}(\rho)}.
\end{equation}
Note that this weight function does not need to be positive and, in general, can have negative values, but its integral over all possible sizes $\xi$ is $1$ by construction. 
If the ACF of $\rho$ is positive, however, $W_l(r)$  can be further identified as the PDF of the size $r$  of correlated structures. 
We can then build the \textit{weighted} average of the size of correlated structures, $\left< l_w \right>$, as :
\begin{equation}
\left< l_w \right> = \int_0^\infty \xi W_l(\xi) \mathrm{d} \xi = \frac{1}{ l_\mathrm{i}(\rho)} \int_0^\infty \Tilde{C}_\rho(\xi) \, \xi  \, \mathrm{d} \xi.
\end{equation}
Then, as was the case for the integral scale,  $l_\mathrm{i}(\rho)$, in many situations 
\begin{equation}
 \int_0^\infty \Tilde{C}_\rho(\xi) \, \xi  \, \mathrm{d} \xi \simeq l_c(\rho)^2,
\end{equation}
which yields:
\begin{empheq}[box=\fbox]{align}
\left< l_w \right> \simeq \frac{ l_c(\rho)^2}{l_{\rm i}(\rho)} \simeq l_c(\rho).
\end{empheq}
Thus, $l_c(\rho)$ measures the average  size of correlated structures, weighted by the correlation coefficients $\Tilde{C}_\rho(\xi)$.  We then call this average size \textit{the average size of the most correlated structures}, 
in order to indicate that it is a weighted average. 

This construction, which relies on the assumption of isotropy, serves to illustrate the physical meaning of $l_c(\rho)$. In the absence of such an assumption, $l_c(\rho)$ is the only quantity that can be defined, 
but can still be  interpreted as a measure of the average size of the most correlated structures. This is in agreement with the picture obtained from Eq.~(\ref{eq:varergodicestimate}) and Eq.~(\ref{eq:varfreqestimate}), 
where the ratio $(R/l_c)^3$ is interpreted as an  effective number of  "independent" samples in the volume $V=(2R)^3$.

\section{Generalised transport equations and conserved quantity} \label{sec:transportacov}
\citet{chandra1951} derived a transport equation for the auto-covariance function $C_\rho$ in a statistically homogeneous isotropic and globally static medium with fixed background density $\overline{\rho}(t)=\rho_0$. We generalize his result to our class of  statistically homogeneous flows that are not necessarily isotropic and with non trivial evolution, i.e.
for which  $\overline{\rho}(t)$ is a function of time and $\overline{\bm{v}} \neq \bm{0}$.

\subsection{Transport equation}

The derivation of the transport equation follows the lines of \citet{chandra1951} but accounting now for the non trivial background flow; it is given in App. \ref{app:transport}.
Expressing everything in terms of the logarithmic density $s$ (see Eq.~(\ref{eq:defs})), we find:
\begin{empheq}[box=\fbox]{align}
0 &=& \frac{\partial}{\partial t} \left( C_{e^s}(\bm{\xi}) \right) + \left( \underline{\underline{\mathrm{L}_V}} (t) \cdot \bm{\xi} \right)^i  \frac{\partial}{\partial \xi_i}  \left(C_{e^s}(\bm{\xi}) \right)  \nonumber \\
&& + \frac{\partial}{\partial \xi_i} \left( R_{e^s, e^s \bm{u}}^i \right)_{\bm{\xi}}  +  \frac{\partial}{\partial \xi_i} \left( R_{e^s, e^s \bm{u}}^i \right)_{-\bm{\xi}}, \label{eq:transportAcov}
\end{empheq}
where $R_{e^s, e^s \bm{u}}^i$ is the cross correlation function of the two fields $e^s$ and $e^s u^i$,  which depends only on the lag $\bm{\xi}$ under the assumption of statistical homogeneity. 
In fact, from the definition of $\bm{u}$, $\overline{e^s u_i}= 0$, so that $R_{e^s, e^s \bm{u}}^i$ is also the cross covariance function of  $e^s$ and $e^s u^i$. 
If one assumes statistical isotropy,
\begin{equation}
R_{e^s, e^s \bm{u}}^i(\bm{\xi})= L_{e^s, e^s u} (|\bm{\xi}|) \, \xi^i.
\end{equation}
Then, the last two terms on the right-hand side of Eq.~(\ref{eq:transportAcov}) can be combined to give $2  \partial_{\xi_i} \left( R_{e^s, e^s \bm{u}}^i \right)_{\bm{\xi}} $  and we recover the result of  \citet{chandra1951}  (his Eq. 13). 
Eq.~(\ref{eq:transportAcov}) thus generalizes the transport equation for the ACF of $\rho$ derived by \citet{chandra1951}   for a non-isotropic, time evolving flow:
\begin{equation}
 \frac{\partial}{\partial t} \left( C_{\rho}(\bm{\xi}) \right) = 2 \partial_{\xi_i} \left( L_{\rho, \rho u} \xi^i \right), \nonumber
\end{equation}
 with the addition of the advection term for relative velocity  $\Delta \overline{v^i} = \overline{v^i}(\bm{x},t) - \overline{v^i}(\bm{x'},t) =    \left( \underline{\underline{\mathrm{L}_V}} (t) \cdot \bm{\xi} \right)^i$, because distortion can only be generated by  the relative motion \citep{kolmogorov1941local,frisch1995}, and without assuming statistical isotropy at all scales. 

As before, we use in the following the two common \textit{physical} assumptions that enforce ergodicity. The covariance and cross covariance functions $C_\rho$ (or $C_{e^s}$) and $R_{e^s, e^s \bm{u}}^i $ 
are both assumed to decay rapidly to $0$ as $|\xi| \rightarrow \infty$ and to be integrable. 

\subsection{Correlation length and conserved quantity} \label{subsec:corelconserved} 

An important quantity characterizing the statistics of the stochastic field $\rho$ (or $e^s$) is the correlation length $l_c(\rho)$ (or $l_c(e^s)$), defined earlier:
\begin{eqnarray}
l_c(\rho)^3 &=& \frac{1}{8  \var{\rho}} \int_{\mathbb{R}^3} C_\rho (\bm{\xi}) \, \mathrm{d} \bm{\xi} \nonumber \\
&=&  \frac{1}{8  \var{e^s}} \int_{\mathbb{R}^3} C_{e^s} (\bm{\xi}) \, \mathrm{d} \bm{\xi} = l_c(e^s)^3. \label{eq:defcorlengt}
\end{eqnarray}
Integrating Eq.~(\ref{eq:transportAcov}) over all possible lags $\bm{\xi}$ yields the conservation equation:
\begin{empheq}[box=\fbox]{align}
 \left(\var{e^s} l_c(e^s)^3 \right)_t \overline{\rho}(t) =   \rm{const} . \label{eq:conservedlcrho}
\end{empheq}
or, in terms of the density field $\rho$:
\begin{eqnarray}
\frac{ \left(\var{\rho} l_c(\rho)^3 \right)_t }{\overline{\rho}(t)} =  \rm{const},
\end{eqnarray}
where quantities of the form $ \left( X \right)_t$ mean that the value of quantity $X$ is taken at time $t$.
These two equations are modified versions of the conservation equation derived by \citet{chandra1951} (his Eq. 17):
\begin{equation}
\int_0^\infty C_\rho(r,t) r^2 \mathrm{d}r =  \rm{const} . \nonumber
\end{equation}
They account for evolution of the average (background) density field and depend explicitly on the correlation length. The detailed derivation of these equations is given in App.~\ref{app:conserved}.

We note that the conserved quantity in Eq.~(\ref{eq:conservedlcrho}) has the dimension of a mass; we will come back to this point later.

\section{Numerical test of the evolution of the correlation length in astrophysical conditions} \label{subsec:numevolvcorrlen}
To test Eq.~\ref{eq:conservedlcrho} in astrophysical conditions, we use the numerical simulations presented in \citet{federrath2012,federrath2013} and used in \citetalias{jaupart2020}. These simulations model the isothermal gravo-turbulent evolution of clouds in periodic boxes of size $L$ with different resolutions $N_{\rm res}$, average density $\rho_0$,  where turbulence is driven at fixed rms Mach numbers $\mathcal{M}$ with solenoidal or compressive forcing or a mixture of both. They belong to the class of statistically homogeneous flows presented in Sec. \ref{subsec:classofflows} where $\bm{V} = \bm{0}$. In each simulation, gravity is added after a gravitationless turbulence  state has developed. As soon as gravity is switched on, the variance of the density field increases due to the condensation of structures. As shown above,  the increase of the variance of $\rho$ is expected to be accompanied by a decrease of the correlation length $l_c(\rho)$. To measure this decrease we use the relation derived in  \citetalias{Jaupart21}: 
\begin{equation}
  \mathrm{Var}\left(\frac{\Sigma}{\esp{ \Sigma }}\right) \simeq \mathrm{Var}\left(\frac{\rho}{\esp{ \rho }}\right) \frac{l_c(\rho)}{R}, \label{eq:relationratiovar}
\end{equation}
that relates the variance of the column density field $\Sigma$ to the variance of $\rho$, $l_c(\rho)$ and the half size of the simulation box $R=L/2$. The derivation of Eq.~(\ref{eq:relationratiovar}) is given in App.~\ref{app:varianceLcRelation}.
Eq.~(\ref{eq:relationratiovar}) thus yields the estimate $\hat{l}_c/R$ of the ratio of the correlation length $l_c(\rho)$ to the half size of the simulation box $R=L/2$:
\begin{equation}
  \frac{\hat{l}_c}{R} \equiv \frac{\mathrm{Var}\left(\frac{\Sigma}{\esp{ \Sigma }}\right) }{\mathrm{Var}\left(\frac{\rho}{\esp{ \rho }}\right) } \simeq  \frac{l_c(\rho)}{R} . \label{eq:lcproxy}
\end{equation}
As shown previously, had Eq.~(\ref{eq:lcproxy}) be an  \textit{exact} equality, we would expect $\hat{l}_c/R \propto \var{\rho}^{-1/3}$  (for fixed $\overline{\rho}$). Eq.~(\ref{eq:lcproxy}), however,  is only a proxy to derive an estimate of $l_c(\rho)$ 
within a factor of order unity which depends on the shape of the auto-covariance function (ACF) of the density field (see Sec. \ref{sec:corrlengthAvSize} and App.~\ref{app:varianceLcRelation}). Furthermore, the ACF is initially that of inertial turbulence and evolves towards an ACF whose shape at short lags is determined by gravity induced dynamics. We  expect the ACF to change with time
between these two regimes. Once the dynamics in high density regions (short scales) starts to be dominated by gravity (the regime we are interested in), 
we expect  the ACF at short lags, while evolving with time, to preserve its functional form. In this regime, i.e. for $\var{\rho}_t \gg \var{\rho}_{t=0}$, we expect   $\hat{l}_c/R \propto \var{\rho}^{-1/3}$ (as mentioned above). 
However, as the simulations can only resolve structures larger than $\Delta x_{\min} = L / N_{\rm res}$, resolution issues can prevent the occurence of this behaviour in the simulations.
Instead, we expect values of $\hat{l}_c/R$ to level off at some point in the simulations.

\begin{figure}
\centering
    \includegraphics[width=\columnwidth]{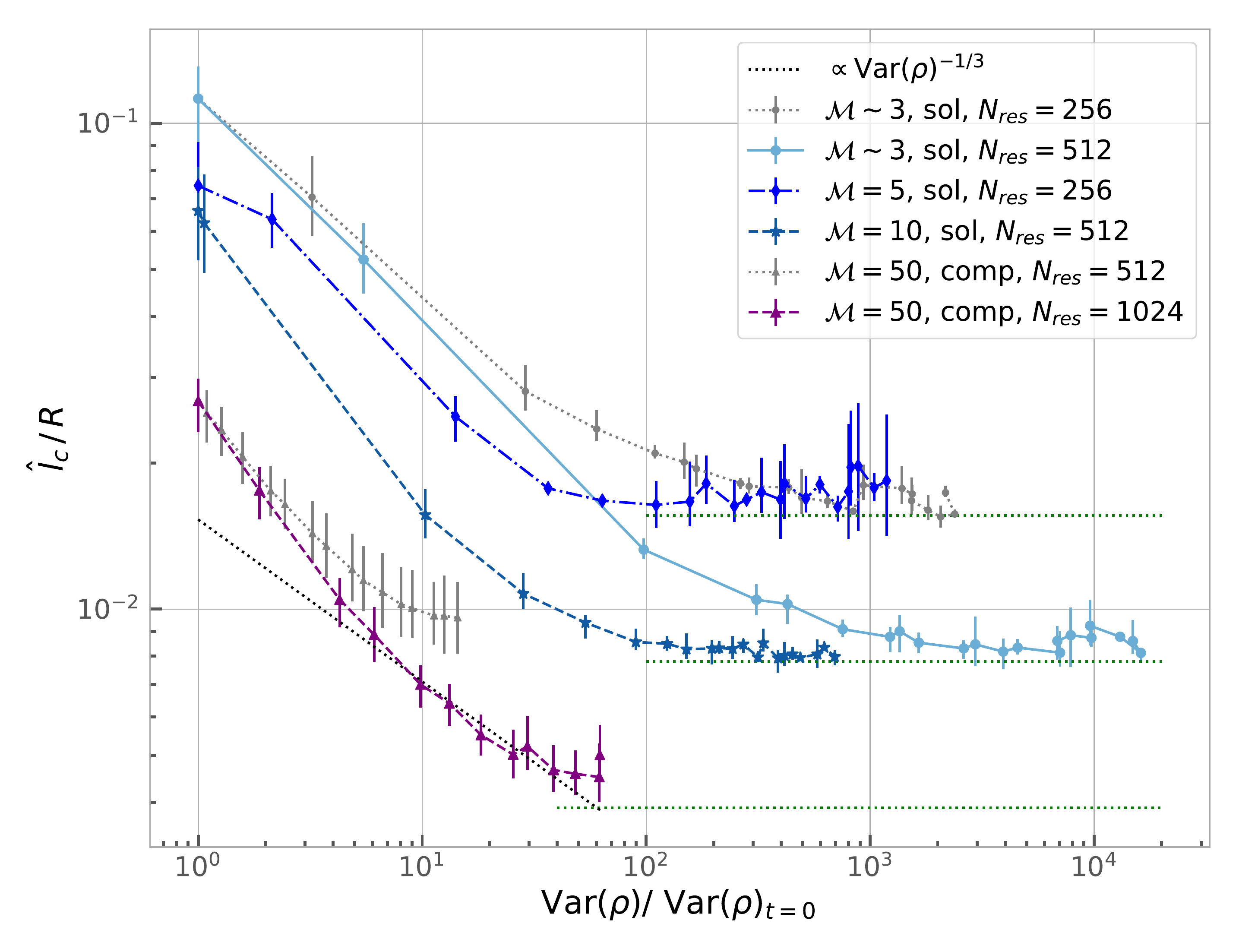}
    \caption{Estimate of correlation length $\hat{l}_c/R$ (Eq.~(\ref{eq:lcproxy})) as a function of ratio  $\var{\rho}_t / \var{\rho}_{t=0}$, for Mach numbers $\mathcal{M} \in \{3, 5, 10, 50\}$  (from light blue to blue, dark blue and purple lines). 
    Two lower resolutions are displayed in grey for $\mathcal{M}=3$ and  $\mathcal{M}=10$, in order to highlight the limitations due to the numerical resolution (as measured by $N_{\rm res}$). 
    The black dotted line corresponds to scaling  $\hat{l}_c/R \propto \var{\rho}^{-1/3}$. Green horizontal dotted lines indicate the value of ratio $ \lambda_J(\rho_{\rm max}) /2 R$ for which $\hat{l}_c/R$ levels off.  }
    \label{fig:num:corr_var}
\end{figure}

Fig.~\ref{fig:num:corr_var} displays estimated values of $\hat{l}_c/R$ as a function of the ratio  $\var{\rho}_t / \var{\rho}_{t=0}$ (which increases with time) from hydrodynamic simulations 
for various Mach number $\mathcal{M}$ and resolution $N_{\rm res}$. As expected, the $\hat{l}_c/R$ ratio decreases as the variance $\var{\rho}$ increases. 
At high variance values (late times), the correlation length is observed to level off at a value that depends on the resolution $N_{\rm res}$, 
corresponding to $\hat{l}_c \simeq 2 \Delta x_{\rm min} = \lambda_J(\rho_{\rm max}) /2$ where 
\begin{equation}
\lambda_J(\rho_{\rm max}) = 4 \Delta x,
\end{equation} 
with $\Delta x = L/N_{\rm res}$ the grid resolution, is the Jeans length at density
\begin{equation}
    \rho_{\rm max} =\frac{\pi c_{\rm s}^2}{16 G \Delta x_{\rm min}^2},
\end{equation}
 with $c_{\rm s}$ the sound speed, above which cloud collapsing features are not resolved \citep{truelove1997}. For simulations at $\mathcal{M} = 50$ at the highest resolution $N_{\rm res} = 1024$, 
we observe that the scaling $\hat{l}_c/R \propto \var{\rho}^{-1/3}$ holds over a decade for $\var{\rho}_t / \var{\rho}_{t=0} \geq 5$. In the other simulations, this scaling law is inhibited by the levelling of $\hat{l}_c$ (save perhaps for the $\mathcal{M} = 10$ one where it holds for half a decade). It would thus be interesting to carry out all simulations with the same highest resolution ($N_{\rm res}=1024$ for example).

The initial values of $\hat{l}_c/R$ yield $\hat{l}_c/L = 0.056 ^{+0.01}_{-0.013}$, $\hat{l}_c/L = 0.037 ^{+0.009}_{-0.006}$, $\hat{l}_c/L = 0.025 ^{+0.005}_{-0.003}$, and $\hat{l}_c/L =0.013 ^{+0.0015}_{-0.0018}$ for $\mathcal{M} = 3,$ $5$, $10$, $50$ respectively. For simulations with  $\mathcal{M} \in \{3, 5, 10\}$, one finds that, within a factor of order unity, $\hat{l}_c \simeq L/\mathcal{M}^2 = \lambda_{\rm s}$, where $\lambda_{\rm s}$ is the sonic length, which is found to be close to the average width of filamentary structures in isothermal turbulence \citep{federrath2016univ}. This is not surprising because $l_c(\rho)$ describes  the average size of the most correlated substructures. 
For the $\mathcal{M} = 50$ simulations, however, $\hat{l}_c$ is about $30$ times larger that $\lambda_{\rm s}=L/2500$. $\lambda_{\rm s}$ is not resolved in these simulations ($N_{\rm res} = 512$ or $1024$), 
which explains the large discrepancy between $\hat{l}_c$ and its expected value $\lambda_{\rm s}$.

The above results show that  Eq.~(\ref{eq:lcproxy}) allows a good approximation of the actual ratio $l_c(\rho) /R$. They also emphasize the fact that correlated substructures are only resolved down to the smallest Jeans length 
that can be achieved in the simulations.  They do not imply that structures larger than $l_c(\rho)$ are not correlated! Such large correlated structures can exist (e.g. large filaments) but they are less correlated than the structures smaller than $l_c(\rho)$ (i.e. they are associated with a lower correlation coefficient $C_\rho/\var{\rho}$). 
Importantly enough, the simulations for the highest resolutions confirm that the quantity $l_c(\rho)^3 \var{e^s}$ is indeed conserved, as expected from our theoretical analysis.

\section{Astrophysical context: star forming clouds and gravity}

Observations and numerical simulations of MCs in dense star-forming regions  have reported a significant  increase of the density variance compared with the one obtained from gravitationless isothermal turbulence (e.g. \citealt{kainulainen2006,kainulainen2009,schneider2012,schneider2013} and references therein for observations and e.g. \citealt{kritsuk2010,ballesteros2011,cho2011,collins2012,federrath2013,lee2015,burkhart2016} for simulations).  Such an increase is believed to be the signature of gravity.

\subsection{Evolution of the correlation length} \label{subsec:evolLc}

In \citetalias{jaupart2020}, we showed that this increase of variance due to gravity  occurs on a short (local)  timescale compared with the  
the typical timescale for variation of the cloud's global mean density $\overline{\rho}$. 


This increase of the variance results in a decrease of the product $\overline{\rho}(t) l_c(e^s)^3$ in order to meet the constraint of the conservation equation (Eq.~(\ref{eq:conservedlcrho})):
\begin{equation}
\var{e^s} \,  \left( 2 l_c(e^s) \right)^3 \, \overline{\rho}(t) = \rm{const}.
\end{equation}
Given the difference of timescales, we can assume, that, during this phase of variance increase, 
the (background) average density  $\overline{\rho}=\mu m_H \bar{n}$ (where $\mu$ and $m_H=1.66\times 10^{-24}$ g denote the mean molecular weight and atomic mass unit, respectively) is almost constant and the conservation equation essentially holds
\begin{empheq}[box=\fbox]{align}
\frac{ \left(l_c(e^s) \right)_t^3}{\left(  l_c(e^s) \right)_{t=t_0}^3}  \simeq  \frac{\left(\var{e^s} \right)_{t=t_0}}{\left(\var{e^s} \right)_{t}} \ll 1 . \label{eq:decreaselc}
\end{empheq}
It is worth stressing that whereas, by construction, $\bar{n}$ is exactly constant in mass conserving simulations, it is not necessarily the case in real star-forming clouds, as it
depends on the bulk flow (see Eq.(\ref{eq::divVhomostrict}) and, e.g., \citealt{robertson2012}).
Thus, the growing impact of gravity on the turbulent flow is accompanied by a drastic decrease of the correlation length of the density field $l_c(e^s) = l_c(\rho)$.

Physically speaking, Eq.~(\ref{eq:decreaselc}) implies that, during the cloud's evolution, the distribution of matter evolves from being concentrated in  weakly correlated structures of average size $\left(  l_c(e^s) \right)_{t=t_0}$ to being concentrated in smaller, denser more and more correlated regions of average size $ \left(l_c(e^s) \right)_t\ll \left(l_c(e^s) \right)_{t=t_0}$. 

This picture is consistent with scenarios of star formation where  the mass concentration in the cloud evolves from large filamentary structures to smaller, denser ones, 
and eventually to small prestellar cores \citep{andre2017interstellar,andre2019role}. 
Within the terminology of the present study, this is described as follows: dense and short scale  tightly correlated substructures ($i.e.$ stellar cores) appear in larger less correlated ones ($i.e.$ filaments). The former ones correspond to objects of average size $l_c(\rho)(t)$ whereas the latter correspond to objects of average (radial) size $l_c(\rho)(t_0)$,
which corresponds to the "initial" correlation length  in early collapsing structures. Indeed, $t_0$ corresponds to the time at which some dense and significant regions within the cloud start to collapse and to deviate from the \textit{global} evolution (contraction or expansion) of the cloud.

It is important to emphasize that the present theoretical framework, which is based on the hypothesis of \textit{statistical} homogeneity, does not rely on any assumption regarding the 
condition or the magnitude of density deviation required for collapse. Furthermore, this framework is able to describe simultaneously a hierarchy of structures spanning a vast range of sizes and densities within the cloud during its evolution.

\subsection{The average mass of  prestellar cores} \label{sec:avCorrMass}
 
 The quantity $\overline{\rho}(t) \var{e^s} \left(2 l_c(\rho) \right)^3 $ has the dimension of a mass (\S\ref{subsec:corelconserved}) and corresponds to the {\it average mass contained in the most correlated structures}, $M_{\rm corr}$:  
 \begin{equation}
M_{\rm corr} \propto \overline{\rho} \var{e^s} \left( 2 l_c(\rho)\right)^3 ,
\end{equation}
with a proportionality coefficient of the order unity that depends on the geometry and where the $2^3$ term stems from the definition of $l_c(\rho)$,  since this latter corresponds to the half size of correlated structures. Initially, $M_{\rm corr}$ is  located within the correlated structures embedded inside large filaments of average width $l_c(\rho)(t_0)$. As collapse proceeds, this (conserved) amount of mass gets distributed in shorter scale, more correlated substructures of average size $l_c(\rho)(t) < l_c(\rho)(t_0) $. 
Eventually, these structures will become prestellar cores. 
Thus, $M_{\rm corr}$ represents ultimately the average mass that is available to form  (prestellar) cores. For a Chabrier like Core Mass Function \citep{chabrier2003galactic,chabrier2005astrophysics}, this {\it average} mass is close to the {\it characteristic} mass.  We calculate below an estimate of its value under typical Milky Way like conditions.


Observations   and theoretical models of star formations indicate that initially, $i.e.$ before the onset of star formation, the  variance characteristic of the PDF of density fluctuations ressembles  that of isothermal  fully developed turbulence  \citep{schneider2013,deoliveira2014,padoan2002,maclow2004,hennebelle2008,hopkins2012,vazquez2019}, i.e. 
\begin{equation}
\var{e^s}(t_0)  = (b\mathcal{M})^2 
\end{equation}
\citep{federrath2008,molina2012,beattie2021}. 
 It remains to determine the correlation length $l_c(\rho)(t_0)$.
While, in case of pure gravitationless turbulence, this latter should be about the sonic length $\lambda_{\rm s}$ , it is not necessarily the case if gravity {\it initially} plays a non negligible role. A detailed determination of the correlation length will be presented in a forthcoming paper. 

Meanwhile, it is safe to take the observed average radial size of correlated filamentary structures, $\sim 0.1$ pc, which is of the order of the sonic length, as an estimate of the "initial" correlation length $l_c(\rho)(t_0)$
(see e.g. \citealt{arzoumanian2011,federrath2016univ,hennebelle2012turbulent,hennebelle2019role} for a more complete discussion). Under typical Milky-Way like conditions, this yields a characteristic correlated mass
\begin{widetext}
\begin{align}
M_{\rm corr} = a_{\rm g} \overline{\rho} \var{e^s}  \, \left( 2 l_c(\rho) \right)^3 = 1.56 M_{\odot}\, \left( \frac{a_{\rm g}}{1} \right) \left(\frac{\mu}{2.0}\right) \left(\frac{\overline{n}}{10^3\text{ cc}^{-1}} \right) \left( \frac{b\, \mathcal{M}}{0.4 \times 5} \right)^2  \left( \frac{l_c(\rho)(t_0)}{0.1 \text{ pc}} \right)^3. \!\!\!\label{eq:ANtypicalMass}
\end{align}
\end{widetext}
It must be kept in mind that the last term of Eq.(\ref{eq:ANtypicalMass}), namely the correlation length, entails a dependence upon the cloud's initial main properties,
average density and Mach number ($l_c(\rho)(t_0)\equiv {l_c}(\rho)_0[{\bar n}, \mathcal{M}]$). 

It must be emphasized that $M_{\rm corr} $  is  set up at the {\it initial} stages of  gravitational contraction, before gravity starts affecting significantly the
properties of turbulence  (see JC20). It corresponds to the mass contained in the most correlated regions embedded in the initial  filamentary structures generated by turbulence. 
It is not necessarily the average mass of all filaments in the cloud.  Similarly, $\overline{n}$ is the initial average density of the cloud, representative of scales large enough that the ergodic estimates 
are accurate. It is not the average density in a (small scale) collapsed subregion.

 
An important property of $M_{\rm corr}$ is that it is not expected to vary significantly among clouds which,  initially, at large scale, meet the typical observed Larson conditions (i.e. $\overline{n}\sim L^{-\eta_d}, \mathcal{M}\sim L^\eta$, with ${\eta_d}\sim 0.7$-1.0, ${\eta}\sim 0.4$-0.5). Indeed, under such conditions, the quantity $\overline{n} \mathcal{M}^2$, thus $M_{\rm corr}$, remains approximately constant.  This remarkable behaviour has been advocated in a different approach, which involves a collapse criterion (namely the virial condition), to explain the apparent universality of the peak of the core mass function for a wide range of stellar cluster conditions \citep{hennebelle2008}. 

As seen from Eq.(\ref{eq:ANtypicalMass}), the theory predicts that, under MW conditions, the average mass available to form prestellar cores,
which is ultimately located in the most correlated structures of size $l_c(\rho)$, is of the order of $\sim 1 M_{\odot}$, in agreement with observations \citep{andre2019role}.

\section{Conclusion}

The theory presented in this Letter, based solely on mass conservation in a {\it statistically} homogeneous medium (not necessarily isotropic nor spatially homogeneous) with non trivial evolution, first provides a description of the ACF and of the evolution of the correlation length $l_c(\rho)$ 
of the density field in star-forming clouds.  We show that this correlation length can be identified as the average size of the most correlated structures (see Sec.~\ref{sec:corrlengthAvSize})
 Then, the theory provides a generalisation of transport equation derived by \citet{chandra1951} for the ACF (Eq.~(\ref{eq:transportAcov})) of density fluctuations in a turbulent medium. It demonstrates the occurence of an invariant in the cloud's evolution,
 which is the average mass contained in the most correlated structures (Eq.~(\ref{eq:conservedlcrho})). For any initial field of density fluctuations this mass is conserved, no matter what dominates the global dynamics (e.g. turbulence or gravity). Comparison with high-resolution numerical simulations (\citet{federrath2012,federrath2013})
 confirms the theoretical relation (Sec.~\ref{subsec:numevolvcorrlen}). 
 This gives an  original and robust description of the physical process occurring in star forming clouds. As collapse progresses within (regions of) the cloud, the variance of the density field increases, so the correlation length $l_c(\rho)$ decreases (Sec.~\ref{subsec:evolLc}), so collapse affects more and more correlated, shorter and shorter scales, yielding the formation of increasingly smaller and clumpier structures. Within this framework, dense and short scale correlated substructures (cores) of average size $l_c(\rho)(t)$ 
form in larger correlated structures (filaments) of average size $l_c(\rho)(t_0) \sim 0.1$ pc. It is worth stressing that the theory, which is based on \textit{statistical} homogeneity, does not constrain fluctuations around the average to be small 
and is able to  simultaneously describe a hierarchy of structures spanning a large range of size and densities  in various environments. 
The theory shows that, under Milky-Way like typical conditions, the invariant average mass contained in the most correlated structures, which will eventually feed (prestellar) cores  is of the order of $\sim 1$ $M_\odot$, providing an appealing explanation for the universality of the peak of the IMF in MW environments.

\appendix

\section{Derivation of the transport equation}\label{app:transport}

Starting from the mass conservation equation (Eq.~(\ref{eq::masscons})) and multiplying it by $\rho' \equiv \rho(\bm{x'})$, one obtains:
\begin{equation}
\rho' \frac{\partial \rho}{\partial t} + \frac{\partial}{\partial x_i} \left( \rho' \rho \, v_i \right) = 0.
\end{equation}
Interchanging the primed and  unprimed quantities in the above equation yields
\begin{equation}
\rho \frac{\partial \rho'}{\partial t} + \frac{\partial}{\partial x'_i} \left( \rho \, \rho' \, v'_i \right) = 0.
\end{equation}
Adding the two equations and taking the average, one obtains \citep{chandra1951}:
\begin{equation}
\frac{\partial}{\partial t}  R_\rho(\bm{x}-\bm{x'},t) +  \frac{\partial}{\partial x_i} \left( \overline{ \rho' \rho \, v_i }\right) + \frac{\partial}{\partial x'_i} \left( \overline{\rho\, \rho' \, v'_i }\right) = 0.
\end{equation}
where
\begin{eqnarray}
R_\rho(\bm{x}-\bm{x'},t) &=& \overline{\rho(\bm{x}) \rho(\bm{x'})}, \nonumber \\
 &=& C_\rho(\bm{x}-\bm{x'},t) +\overline{\rho}(t)^2,
\end{eqnarray}
is the correlation function.

Decomposing $\bm{v} $ into the mean velocity $\bm{V}$ and turbulent component $\bm{u}$ ($\bm{v} = \bm{V} + \bm{u}$), we obtain:
\begin{eqnarray}
0 &=& \frac{\partial}{\partial t}  C_\rho(\bm{\xi},t) - 2  \,C_\rho(\bm{\xi}) \frac{1}{\overline{\rho}} \frac{\partial}{\partial t} \overline{\rho} + (V_i - V'_i)  \frac{\partial}{\partial \xi_i} C_\rho(\bm{\xi}) \nonumber \\
&&+ \frac{\partial}{\partial x_i} \left( \overline{ \rho' \rho \, u_i }\right) + \frac{\partial}{\partial x'_i} \left( \overline{\rho\, \rho' \, u'_i }\right),
\end{eqnarray}
where $\xi = \bm{x}-\bm{x'}$ and where we have used Eq.~(\ref{eq::divVhomostrict}). Then, dividing both sides by $\overline{\rho}(t)^2$ and using Eq.~(\ref{eq:generalformV}), we obtain:
\begin{eqnarray}
0 &=& \frac{\partial}{\partial t} \left( \frac{C_\rho(\bm{\xi})}{\overline{\rho}^2} \right) + \left( \underline{\underline{\mathrm{L}_V}} (t) \cdot \bm{\xi} \right)^i  \frac{\partial}{\partial \xi_i}  \left( \frac{C_\rho(\bm{\xi})}{\overline{\rho}^2} \right) \nonumber \\
&& + \frac{\partial}{\partial x_i} \left( \overline{ \frac{\rho' \rho}{\overline{\rho}^2} \, u_i }\right) + \frac{\partial}{\partial x'_i} \left( \overline{\frac{\rho\, \rho'}{\overline{\rho}^2} \, u'_i }\right).
\end{eqnarray}
Expressing everything in terms of the logarithmic density $s$ (see Eq.~(\ref{eq:defs})), we find:
\begin{empheq}[box=\fbox]{align}
0 &=& \frac{\partial}{\partial t} \left( C_{e^s}(\bm{\xi}) \right) + \left( \underline{\underline{\mathrm{L}_V}} (t) \cdot \bm{\xi} \right)^i  \frac{\partial}{\partial \xi_i}  \left(C_{e^s}(\bm{\xi}) \right)  \nonumber \\
&& + \frac{\partial}{\partial x_i} \left( \overline{ e^{s'} \, e^{s} \, u_i }\right) + \frac{\partial}{\partial x'_i} \left( \overline{ e^{s} \, e^{s'} \, u'_i }\right), \\
&=& \frac{\partial}{\partial t} \left( C_{e^s}(\bm{\xi}) \right) + \left( \underline{\underline{\mathrm{L}_V}} (t) \cdot \bm{\xi} \right)^i  \frac{\partial}{\partial \xi_i}  \left(C_{e^s}(\bm{\xi}) \right)  \nonumber \\
&& + \frac{\partial}{\partial \xi_i} \left( R_{e^s, e^s \bm{u}}^i \right)_{\bm{\xi}}  +  \frac{\partial}{\partial \xi_i} \left( R_{e^s, e^s \bm{u}}^i \right)_{-\bm{\xi}}, \label{app:eq:transportAcov}
\end{empheq}
where $R_{e^s, e^s \bm{u}}^i$ is the cross correlation function of the two fields $e^s$ and $e^s u^i$,  which depends only on the lag $\bm{\xi}= \bm{x}-\bm{x'}$ under the assumption of statistical homogeneity.

\section{Conserved quantity} \label{app:conserved}
To obtain the conserved quantity Eq.~(\ref{eq:conservedlcrho}), one starts by noting that
\begin{eqnarray}
\frac{\mathrm{d}}{\mathrm{d} t} \left( \iiint_{\mathbb{R}^3} C_{e^s} (\bm{\xi}) \, \mathrm{d} \bm{\xi} \right) &= &\iiint_{\mathbb{R}^3}  \partial_t C_{e^s} (\bm{\xi},t) \, \mathrm{d} \bm{\xi}  \nonumber \\
&=& \! - \! \iiint_{\mathbb{R}^3}  \! \! \left( \underline{\underline{\mathrm{L}_V}} (t) \! \cdot \! \bm{\xi} \right)^i  \! \frac{\partial}{\partial \xi_i}  \left(C_{e^s}(\bm{\xi}) \right) \, \mathrm{d} \bm{\xi} \nonumber \\
&&- 2\! \iint_{ \mathrm{"} \partial \mathbb{R}^3 \mathrm{"}} \! \!  \!  \! R_{e^s, e^s \bm{u}}^i (\bm{\xi}) \, \mathrm{d} S^i ,
\end{eqnarray}
where the surface integral (the second term on the right hand side of the equation) vanishes due to the assumption on $R_{e^s, e^s \bm{u}}^i $. The first term on the right hand side can be rewritten such that:
\begin{eqnarray}
\frac{\mathrm{d}}{\mathrm{d} t} \left( \iiint_{\mathbb{R}^3} C_{e^s} (\bm{\xi}) \, \mathrm{d} \bm{\xi} \right) &=& \! - \! \iiint_{\mathbb{R}^3}  \! \bm{\nabla} \! \cdot \! \left( C_{e^s}(\bm{\xi}) \,  \underline{\underline{\mathrm{L}_V}} (t) \! \cdot \! \bm{\xi}  \right) \, \mathrm{d} \bm{\xi} \nonumber \\
&&+  \iiint_{\mathbb{R}^3} (\bm{\nabla} \! \cdot \! \bm{V}) C_{e^s}(\bm{\xi})\, \mathrm{d} \bm{\xi} .
\end{eqnarray}
The first term on the right hand side can be turned into a surface integral, which also vanishes due to the assumption on $C_{e^s}$. We are thus left with:
\begin{equation}
\frac{\mathrm{d}}{\mathrm{d} t} \left( \iiint_{\mathbb{R}^3} C_{e^s} (\bm{\xi}) \, \mathrm{d} \bm{\xi} \right)  = -  \frac{\mathrm{d} \mathrm{ln}(\overline{\rho})}{\mathrm{d} t} \times \left( 8  \var{e^s} l_c(e^s)^3 \right), 
\end{equation}
which yields:
\begin{equation}
\frac{\mathrm{d}}{\mathrm{d} t} \left(\var{e^s} l_c(e^s)^3 \right) =  - \left(\var{e^s} l_c(e^s)^3 \right) \times \frac{\mathrm{d} \mathrm{ln}(\overline{\rho})}{\mathrm{d} t}, \label{app:eq:conservedlcs}
\end{equation}
and
\begin{empheq}[box=\fbox]{align}
 \left(\var{e^s} l_c(e^s)^3 \right)_t \overline{\rho}(t) =   \rm{const} . \label{app:eq:conservedlcrho}
\end{empheq}

In principle, the integral in Eq.~(\ref{eq:defcorlengt}) must be carried out over all possible lags $\bm{\xi}$ and hence over the whole space $\mathbb{R}^3$, 
which may seem conceptually problematic as we want to deal with a cloud of finite size. As regards the bulk flow, however, we rely on the same line of reasoning as in statistical mechanics:  
if the actual subspace of permitted lags is large enough, it can be assimilated to the whole space $\mathbb{R}^3$. The argument is the following. If $\Omega$, the subspace of permitted lags, 
is such that its volume $|\Omega|$ is $ \gg l_c(e^s)^3 $, i.e. contains a large number of correlation volumes, and if  $C_\rho$ (or $C_{e^s}$) tends to $0$ as $|\xi| \rightarrow \infty$  and is integrable, 
the integral over $\Omega$ can be seen as an integral over $\mathbb{R}^3$. 

To understand the meaning of the conserved quantity in Eq.~(\ref{eq:conservedlcrho}) (or Eq.(\ref{app:eq:conservedlcrho})) and the approximation made, we now consider a finite subspace of permitted lags. 
Let $\Omega_t$ be the "average" volume of space describing the cloud under study, evolving with the average velocity field 
$\overline{\bm{v}} = \bm{V}(\bm{x},t) + \overline{\bm{u}}(t)=  \underline{\underline{\mathrm{L}_V}} (t) \cdot \bm{x} + \bm{c}_V(t) + \overline{\bm{u}}(t)$. $\Omega_t$ is hence a mass conserving domain 
and, like $\overline{\rho}(t)$, is allowed to evolve with time.  If $\Omega_t$ possesses point symmetry, then  the subspace of permitted lags is simply $\Omega_{t,\xi} = 2 \, \Omega_t$.  
This subspace is evolving with the \textit{relative} velocity field $\Delta \overline{\bm{v}} = \overline{\bm{v}}(\bm{x},t) - \overline{\bm{v}}(\bm{x'},t) =   \underline{\underline{\mathrm{L}_V}} (t) \cdot \bm{\xi}$, 
because distorsion can only be generated by  the relative motion \citep{kolmogorov1941local,frisch1995}. Due to Reynolds' transport theorem, one has:
%
\begin{align}
\frac{\mathrm{d}}{\mathrm{d} t} \left( \frac{1}{|\Omega_t|} \iiint_{\Omega_{t, \bm{\xi}}}  \!\!\! C_{e^s} (\bm{\xi}) \, \mathrm{d} \bm{\xi} \right) &= & - \frac{1}{|\Omega_t|^2}\frac{\mathrm{d} |\Omega_t|}{\mathrm{d} t}  \iiint_{\Omega_{t, \bm{\xi}}} \!\!\! C_{e^s} (\bm{\xi}) \, \mathrm{d} \bm{\xi}  + \frac{1}{|\Omega_t|}  \iiint_{\Omega_{t, \bm{\xi}}}  \!\!\! C_{e^s} (\bm{\xi}) \times \left( \bm{\nabla} \! \cdot \,  \Delta \overline{\bm{v}} \right) \, \mathrm{d} \bm{\xi}   \nonumber \\
&& +  \frac{1}{|\Omega_t|} \iiint_{\Omega_{t, \bm{\xi}}} \!\!\! \left( \partial_t + ( \underline{\underline{\mathrm{L}_V}} (t) \! \cdot \! \bm{\xi} ) \cdot \bm{\nabla}   \right) \! C_{e^s} (\bm{\xi}) \, \mathrm{d} \bm{\xi},
\end{align}
where:
\begin{align}
 \bm{\nabla} \! \cdot \,  \Delta \overline{\bm{v}}  =  \bm{\nabla} \! \cdot \, \left( \underline{\underline{\mathrm{L}_V}} (t) \cdot \bm{\xi} \right) =  \bm{\nabla} \! \cdot \,  \bm{V} =  - \frac{\mathrm{d} \mathrm{ln}(\overline{\rho})}{\mathrm{d} t}, \\
 \frac{1}{|\Omega_t|}\frac{\mathrm{d} |\Omega_t|}{\mathrm{d} t}  =   - \frac{\mathrm{d} \mathrm{ln}(\overline{\rho})}{\mathrm{d} t} = \bm{\nabla} \! \cdot \,  \bm{V} .
\end{align}
This leads to:
\begin{eqnarray}
\frac{\mathrm{d}}{\mathrm{d} t} \left( \frac{1}{|\Omega_t|} \iiint_{\Omega_{t, \bm{\xi}}} C_{e^s} (\bm{\xi}) \, \mathrm{d} \bm{\xi} \right) &= &  \frac{1}{|\Omega_t|} \iiint_{\Omega_{t, \bm{\xi}}} \!\!\! \left( \partial_t + ( \underline{\underline{\mathrm{L}_V}} (t) \! \cdot \! \bm{\xi} ) \cdot \bm{\nabla}   \right) \! C_{e^s} (\bm{\xi}) \, \mathrm{d} \bm{\xi} = -  2  \frac{1}{|\Omega_t|} \! \iint_{ \partial \Omega_{t, \bm{\xi}}} \! \!  \!  \! R_{e^s, e^s \bm{u}}^i (\bm{\xi}) \, \mathrm{d} S^i. \label{app:eq:conservlcfinitedomain}
\end{eqnarray}
Assuming now that the contribution from the surface integral at the r.h.s of Eq.~(\ref{app:eq:conservlcfinitedomain}) is negligible, 
$i.e.$ assuming that $R_{e^s, e^s \bm{u}}^i $ decays rapidly to $0$ at large lags $\bm{\xi}$ and that $\Omega_t$ (and hence $\Omega_{t, \bm{\xi}}$) is large enough (for example such that  $|\Omega_t| \gg l_c(e^s)^3 $), we are left with:
\begin{eqnarray}
 \frac{1}{|\Omega_t|} \iiint_{\Omega_{t, \bm{\xi}}} C_{e^s} (\bm{\xi}) \, \mathrm{d} \bm{\xi}  \simeq 8 \, \var{e^s} \frac{l_c(e^s)^3}{|\Omega_t|} = \mathrm{const} .
\end{eqnarray}
Using the fact that $ \overline{\rho}(t) |\Omega_t| = M(\Omega_t)= \mathrm{const}$, we obtain:
\begin{empheq}[box=\fbox]{align}
 \var{e^s} \, l_c(e^s)^3 \, \overline{\rho}(t) = \mathrm{const}, \label{app:eq:conservedlcsOmega}
\end{empheq}
which is Eq.~(\ref{app:eq:conservedlcrho}) (or Eq.~(\ref{eq:conservedlcrho})). These calculations are valid for any (mass conserving) sub-domain $\Omega_t$ that is large enough for the surface integral on the r.h.s of Eq.~(\ref{app:eq:conservlcfinitedomain}) to be negligible. Eq.~(\ref{app:eq:conservedlcsOmega})  therefore implies that the fundamental quantity $ \var{e^s} \, l_c(e^s)^3 \, \overline{\rho}(t)$ is conserved. 

\section{Estimate of the correlation length from the ratio of column density to volume density variances.} \label{app:varianceLcRelation}
We give the derivation of Eq.~(\ref{app:eq:relationratiovar}). For a cubic simulation domain of size $L$, projecting the density field along one of the 3 principal directions of the cube leads to a statistically homogeneous column density field such that : 
\begin{equation}
 \esp{\Sigma(x,y)} = \esp{\rho} \times L. 
 \end{equation}
In a cubic box, the ACF of $\Sigma$ is
\begin{eqnarray}
    C_\Sigma(\bm{r}) &=& \esp{\left(\Sigma(\bm{u} + \bm{r})- \esp{\rho} \, L\right) \left(\Sigma(\bm{u})- \esp{\rho} \, L\right) } \nonumber\\
    &=& \int_{[-L/2,L/2]^2} C_\rho(\bm{r}, z-z')\, \mathrm{d} z \, \mathrm{d}z' \nonumber \\
    &=& \int_{[-L,L]} C_\rho(\bm{r},u) \mathrm{d}u \int_{-L+|u|}^{L-|u|} \frac{\mathrm{d} v}{2} \nonumber \\
    &=& L \int_{[-L,L]} C_\rho(\bm{r},u) \left(1-\frac{|u|}{L} \right) \mathrm{d}u,
\end{eqnarray}
while the variance is 
\begin{eqnarray}
\mathrm{Var}(\Sigma)&=&C_\Sigma(\bm{0}) = L \int_{[-L,L]} C_\rho(\bm{0},u)  \left(1-\frac{|u|}{L} \right) \mathrm{d}u.
\end{eqnarray}
Thus, assuming that the density field is statistically isotropic at small scales (i.e. the ACF is isotropic at short lags), one obtains:
\begin{eqnarray}
\mathrm{Var}(\Sigma) &\simeq& L \int_{[-L,L]} C_\rho(|u|)  \left(1-\frac{|u|}{L} \right) \mathrm{d}u. \label{eq:varsigmaC}
\end{eqnarray}
Provided  the correlation length of the density field is much smaller than the size of the box $L$ ($i.e.$ $l_c(\rho)\ll L$), one can approximate the integral on the r.h.s of Eq.~(\ref{eq:varsigmaC}) by the following expression:
\begin{equation}
 \int_{[-L,L]} C_\rho(|u|)  \left(1-\frac{|u|}{L} \right) \mathrm{d}u \simeq 2 l_{\mathrm i}(\rho) \mathrm{Var}(\rho) \simeq 2 l_c(\rho)  \mathrm{Var}(\rho),
\end{equation}
where $ l_{\mathrm i}(\rho)$ is the integral scale of the density field. Thus, 
\begin{equation}
\mathrm{Var}(\Sigma) \simeq 2\,L\, l_c(\rho)\, \mathrm{Var}(\rho). \label{eq:variancesigma}
\end{equation}
This yields
\begin{empheq}[box=\fbox]{align}
   \mathrm{Var}\left(\frac{\Sigma}{\esp{ \Sigma }}\right) \simeq \mathrm{Var}\left(\frac{\rho}{\esp{ \rho }}\right) \frac{2\,l_c(\rho)}{L} = \mathrm{Var}\left(\frac{\rho}{\esp{ \rho }}\right) \frac{l_c(\rho)}{R}, \label{app:eq:relationratiovar}
\end{empheq}
where $R=L/2$. This is an important result because it provides a {\it measure of $l_c(\rho)/R$ independently of the ACF}.

\bibliography{biblio_these.bib}{}

\begin{thebibliography}{}
\expandafter\ifx\csname natexlab\endcsname\relax\def\natexlab#1{#1}\fi
\providecommand{\url}[1]{\href{#1}{#1}}
\providecommand{\dodoi}[1]{doi:~\href{http://doi.org/#1}{\nolinkurl{#1}}}
\providecommand{\doeprint}[1]{\href{http://ascl.net/#1}{\nolinkurl{http://ascl.net/#1}}}
\providecommand{\doarXiv}[1]{\href{https://arxiv.org/abs/#1}{\nolinkurl{https://arxiv.org/abs/#1}}}

\bibitem[{Andr{\'e}(2017)}]{andre2017interstellar}
Andr{\'e}, P. 2017, Comptes Rendus Geoscience, 349, 187

\bibitem[{Andr{\'e} {et~al.}(2019)Andr{\'e}, Arzoumanian, K{\"o}nyves,
  Shimajiri, \& Palmeirim}]{andre2019role}
Andr{\'e}, P., Arzoumanian, D., K{\"o}nyves, V., Shimajiri, Y., \& Palmeirim,
  P. 2019, Astronomy \& Astrophysics, 629, L4

\bibitem[{Arzoumanian {et~al.}(2011)Arzoumanian, Andr{\'e}, Didelon,
  K{\"o}nyves, Schneider, Men’shchikov, Sousbie, Zavagno, Bontemps,
  Di~Francesco, {et~al.}}]{arzoumanian2011}
Arzoumanian, D., Andr{\'e}, P., Didelon, P., {et~al.} 2011, Astronomy \&
  Astrophysics, 529, L6

\bibitem[{Ballesteros-Paredes {et~al.}(2011)Ballesteros-Paredes,
  V{\'a}zquez-Semadeni, Gazol, Hartmann, Heitsch, \&
  Col{\'\i}n}]{ballesteros2011}
Ballesteros-Paredes, J., V{\'a}zquez-Semadeni, E., Gazol, A., {et~al.} 2011,
  Monthly Notices of the Royal Astronomical Society, 416, 1436

\bibitem[{Batchelor(1953)}]{batchelor1953}
Batchelor, G.~K. 1953, The theory of homogeneous turbulence (Cambridge
  university press)

\bibitem[{Beattie {et~al.}(2021)Beattie, Mocz, Federrath, \&
  Klessen}]{beattie2021}
Beattie, J.~R., Mocz, P., Federrath, C., \& Klessen, R.~S. 2021, Monthly
  Notices of the Royal Astronomical Society, 504, 4354

\bibitem[{{Buchert} \& {Ehlers}(1997)}]{Buchert1997}
{Buchert}, T., \& {Ehlers}, J. 1997, Astronomy \& Astrophysics, 320, 1.
\newblock \doarXiv{astro-ph/9510056}

\bibitem[{Burkhart {et~al.}(2016)Burkhart, Stalpes, \& Collins}]{burkhart2016}
Burkhart, B., Stalpes, K., \& Collins, D.~C. 2016, The Astrophysical Journal
  Letters, 834, L1

\bibitem[{Chabrier(2003)}]{chabrier2003galactic}
Chabrier, G. 2003, PASP, 115, 763

\bibitem[{Chabrier(2005)}]{chabrier2005astrophysics}
---. 2005, Astrophysics and Space Science Library, Vol. 327, The Initial Mass
  Function 50 Years Later,  Springer

\bibitem[{Chandrasekhar(1951{\natexlab{a}})}]{chandra1951}
Chandrasekhar, S. 1951{\natexlab{a}}, Proceedings of the Royal Society of
  London. Series A. Mathematical and Physical Sciences, 210, 18

\bibitem[{Chandrasekhar(1951{\natexlab{b}})}]{chandra1951gravity}
---. 1951{\natexlab{b}}, Proceedings of the Royal Society of London. Series A.
  Mathematical and Physical Sciences, 210, 26

\bibitem[{Cho \& Kim(2011)}]{cho2011}
Cho, W., \& Kim, J. 2011, Monthly Notices of the Royal Astronomical Society:
  Letters, 410, L8

\bibitem[{Collins {et~al.}(2012)Collins, Kritsuk, Padoan, Li, Xu, Ustyugov, \&
  Norman}]{collins2012}
Collins, D.~C., Kritsuk, A.~G., Padoan, P., {et~al.} 2012, The Astrophysical
  Journal, 750, 13

\bibitem[{De~Oliveira {et~al.}(2014)De~Oliveira, Schneider, Mer{\'\i}n, Prusti,
  Ribas, Cox, Vavrek, K{\"o}nyves, Arzoumanian, Puga,
  {et~al.}}]{deoliveira2014}
De~Oliveira, C.~A., Schneider, N., Mer{\'\i}n, B., {et~al.} 2014, Astronomy \&
  Astrophysics, 568, A98

\bibitem[{Federrath(2016)}]{federrath2016univ}
Federrath, C. 2016, Monthly Notices of the Royal Astronomical Society, 457, 375

\bibitem[{Federrath \& Klessen(2012)}]{federrath2012}
Federrath, C., \& Klessen, R.~S. 2012, The Astrophysical Journal, 761, 156,
  \dodoi{10.1088/0004-637X/761/2/156}

\bibitem[{Federrath \& Klessen(2013)}]{federrath2013}
---. 2013, The Astrophysical Journal, 763, 51

\bibitem[{Federrath {et~al.}(2008)Federrath, Klessen, \&
  Schmidt}]{federrath2008}
Federrath, C., Klessen, R.~S., \& Schmidt, W. 2008, The Astrophysical Journal
  Letters, 688, L79

\bibitem[{Frisch(1995)}]{frisch1995}
Frisch, U. 1995, Turbulence: the legacy of AN Kolmogorov (Cambridge university
  press)

\bibitem[{Heinesen(2020)}]{heinesen2020}
Heinesen, A. 2020, Journal of Cosmology and Astroparticle Physics, 2020, 052

\bibitem[{Hennebelle \& Chabrier(2008)}]{hennebelle2008}
Hennebelle, P., \& Chabrier, G. 2008, The Astrophysical Journal, 684, 395

\bibitem[{Hennebelle \& Falgarone(2012)}]{hennebelle2012turbulent}
Hennebelle, P., \& Falgarone, E. 2012, The Astronomy and Astrophysics Review,
  20, 55

\bibitem[{Hennebelle \& Inutsuka(2019)}]{hennebelle2019role}
Hennebelle, P., \& Inutsuka, S.-i. 2019, Frontiers in Astronomy and Space
  Sciences, 6, 5

\bibitem[{Hopkins(2012)}]{hopkins2012}
Hopkins, P.~F. 2012, Monthly Notices of the Royal Astronomical Society, 423,
  2037

\bibitem[{Jaupart \& Chabrier(2020)}]{jaupart2020}
Jaupart, E., \& Chabrier, G. 2020, The Astrophysical Journal Letters, 903, L2

\bibitem[{Jaupart \& Chabrier(2021)}]{Jaupart21}
---. 2021, submitted in A\&A, accepted under revisions

\bibitem[{Kainulainen {et~al.}(2009)Kainulainen, Beuther, Henning, \&
  Plume}]{kainulainen2009}
Kainulainen, J., Beuther, H., Henning, T., \& Plume, R. 2009, Astronomy \&
  Astrophysics, 508, L35

\bibitem[{Kainulainen {et~al.}(2006)Kainulainen, Lehtinen, \&
  Harju}]{kainulainen2006}
Kainulainen, J., Lehtinen, K., \& Harju, J. 2006, Astronomy \& Astrophysics,
  447, 597

\bibitem[{Kolmogorov(1941)}]{kolmogorov1941local}
Kolmogorov, A.~N. 1941, Cr Acad. Sci. URSS, 30, 301

\bibitem[{Kritsuk {et~al.}(2010)Kritsuk, Norman, \& Wagner}]{kritsuk2010}
Kritsuk, A.~G., Norman, M.~L., \& Wagner, R. 2010, The Astrophysical Journal
  Letters, 727, L20

\bibitem[{Ledoux \& Walraven(1958)}]{ledoux1958}
Ledoux, P., \& Walraven, T. 1958, in Astrophysics II: Stellar
  Structure/Astrophysik II: Sternaufbau (Springer), 353--604

\bibitem[{Lee {et~al.}(2015)Lee, Chang, \& Murray}]{lee2015}
Lee, E.~J., Chang, P., \& Murray, N. 2015, The Astrophysical Journal, 800, 49

\bibitem[{Mac~Low \& Klessen(2004)}]{maclow2004}
Mac~Low, M.-M., \& Klessen, R.~S. 2004, Reviews of modern physics, 76, 125

\bibitem[{Molina {et~al.}(2012)Molina, Glover, Federrath, \&
  Klessen}]{molina2012}
Molina, F., Glover, S.~C., Federrath, C., \& Klessen, R.~S. 2012, Monthly
  Notices of the Royal Astronomical Society, 423, 2680

\bibitem[{Padoan \& Nordlund(2002)}]{padoan2002}
Padoan, P., \& Nordlund, {\AA}. 2002, The Astrophysical Journal, 576, 870

\bibitem[{Pan {et~al.}(2018)Pan, Padoan, \& Nordlund}]{pan2018}
Pan, L., Padoan, P., \& Nordlund, {\AA}. 2018, The Astrophysical Journal
  Letters, 866, L17

\bibitem[{Pan {et~al.}(2019{\natexlab{a}})Pan, Padoan, \& Nordlund}]{Pan2019A}
---. 2019{\natexlab{a}}, The Astrophysical Journal, 876, 90

\bibitem[{Pan {et~al.}(2019{\natexlab{b}})Pan, Padoan, \& Nordlund}]{Pan2019B}
---. 2019{\natexlab{b}}, The Astrophysical Journal, 881, 155

\bibitem[{Papoulis \& Pillai(1965)}]{papoulis1965}
Papoulis, A., \& Pillai, S. 1965, Variables Stochastic Processes. Mc Graw
  McGraw-Hill, New York, NY

\bibitem[{{Peebles}(1973)}]{Peebles1973}
{Peebles}, P.~J.~E. 1973, The Astrophysical Journal, 185, 413,
  \dodoi{10.1086/152431}

\bibitem[{Penrose(1979)}]{penrose1979foundations}
Penrose, O. 1979, Reports on Progress in Physics, 42, 1937

\bibitem[{Pope(1985)}]{pope1985}
Pope, S.~B. 1985, Progress in energy and combustion science, 11, 119

\bibitem[{Robertson \& Goldreich(2012)}]{robertson2012}
Robertson, B., \& Goldreich, P. 2012, The Astrophysical Journal Letters, 750,
  L31

\bibitem[{Schneider {et~al.}(2013)Schneider, Andr{\'e}, K{\"o}nyves, Bontemps,
  Motte, Federrath, Ward-Thompson, Arzoumanian, Benedettini, Bressert,
  {et~al.}}]{schneider2013}
Schneider, N., Andr{\'e}, P., K{\"o}nyves, V., {et~al.} 2013, The Astrophysical
  Journal Letters, 766, L17

\bibitem[{Schneider {et~al.}(2012)Schneider, Csengeri, Hennemann, Motte,
  Didelon, Federrath, Bontemps, Di~Francesco, Arzoumanian, Minier,
  {et~al.}}]{schneider2012}
Schneider, N. e.~a., Csengeri, T., Hennemann, M., {et~al.} 2012, Astronomy \&
  Astrophysics, 540, L11

\bibitem[{Truelove {et~al.}(1997)Truelove, Klein, McKee, Holliman~II, Howell,
  \& Greenough}]{truelove1997}
Truelove, J.~K., Klein, R.~I., McKee, C.~F., {et~al.} 1997, The Astrophysical
  Journal Letters, 489, L179

\bibitem[{V{\'a}zquez-Semadeni {et~al.}(2019)V{\'a}zquez-Semadeni, Palau,
  Ballesteros-Paredes, G{\'o}mez, \& Zamora-Avil{\'e}s}]{vazquez2019}
V{\'a}zquez-Semadeni, E., Palau, A., Ballesteros-Paredes, J., G{\'o}mez, G.~C.,
  \& Zamora-Avil{\'e}s, M. 2019, Monthly Notices of the Royal Astronomical
  Society, 490, 3061

\bibitem[{Vigneron(2021)}]{vigneron2021}
Vigneron, Q. 2021, Physical Review D, 103, 064064

\end{thebibliography}
\bibliographystyle{aasjournal}



\end{document}